\documentclass[twocolumn,prb,showpacs,superscriptaddress,preprintnumbers,amsmath,amssymb]{revtex4}

\usepackage[dvips]{graphicx}
\usepackage{dcolumn}
\usepackage{bm}
\begin{document}
\title{Electronic structure of $RE$AuMg and $RE$AgMg ($RE$ = Eu, Gd, Yb)}

\author{Jan Gegner}
  \affiliation{II. Physikalisches Institut, Universit\"{a}t zu K\"{o}ln,
   Z\"{u}lpicher Strasse 77, D-50937 K\"{o}ln, Germany}
\author{T. C. Koethe}
  \affiliation{II. Physikalisches Institut, Universit\"{a}t zu K\"{o}ln,
   Z\"{u}lpicher Strasse 77, D-50937 K\"{o}ln, Germany}
\author{Hua Wu}
  \affiliation{II. Physikalisches Institut, Universit\"{a}t zu K\"{o}ln,
   Z\"{u}lpicher Strasse 77, D-50937 K\"{o}ln, Germany}
\author{H. Hartmann}
  \affiliation{II. Physikalisches Institut, Universit\"{a}t zu K\"{o}ln,
   Z\"{u}lpicher Strasse 77, D-50937 K\"{o}ln, Germany}
\author{T. Lorenz}
  \affiliation{II. Physikalisches Institut, Universit\"{a}t zu K\"{o}ln,
   Z\"{u}lpicher Strasse 77, D-50937 K\"{o}ln, Germany}
\author{T. Fickenscher}
  \affiliation{Institut f\"{u}r Anorganische und Analytische Chemie,
   Westf\"{a}lische Wilhelms-Universit\"{a}t M\"{u}nster,
   Corrensstrasse 30, D-48149 M\"{u}nster, Germany}
\author{R. P\"{o}ttgen}
  \affiliation{Institut f\"{u}r Anorganische und Analytische Chemie,
   Westf\"{a}lische Wilhelms-Universit\"{a}t M\"{u}nster,
   Corrensstrasse 30, D-48149 M\"{u}nster, Germany}
\author{L. H. Tjeng}
  \affiliation{II. Physikalisches Institut, Universit\"{a}t zu K\"{o}ln,
   Z\"{u}lpicher Strasse 77, D-50937 K\"{o}ln, Germany}

\date{\today}

\hyphenation{TiNiSi CePdMg ZrNiAl CePtMg CeAuMg GdAuMg GdAgMg YbAuMg EuAuMg
pho-to-emis-sion Lan-than-oid-spe-zi-fi-sche}

\begin{abstract}

We have investigated the electronic structure of the equiatomic
EuAuMg, GdAuMg, YbAuMg and GdAgMg intermetallics using x-ray
photoelectron spectroscopy. The spectra revealed that the Yb and
Eu are divalent while the Gd is trivalent. The spectral weight in
the vicinity of the Fermi level is dominated by the mix of Mg
$s$, Au/Ag $sp$ and $RE$ $spd$ bands, and not by the $RE$ $4f$.
We also found that the Au and Ag $d$ bands are extraordinarily
narrow, as if the noble metal atoms were impurities submerged in
a low density $sp$ metal host. The experimental results were
compared with band structure calculations, and we found good
agreement provided that the spin-orbit interaction in the Au an
Ag $d$ bands is included and correlation effects in an open $4f$
shell are accounted for using the local density approximation +
Hubbard $U$ scheme. Nevertheless, limitations of such a
mean-field scheme to explain excitation spectra are also evident.

\end{abstract}


\maketitle

Many equiatomic rare-earth ($RE$) transition-metal ($T$) magnesium intermetallics have
been synthesized in the last decade.\cite{P1,P2,P3,P4,P5,P6,P7,P8,P9,P10,P11,P12} $RET$Mg
materials with Yb and Eu as rare-earth elements crystallize in the orthorhombic TiNiSi
structure,\cite{P16} while those with the other rare-earths adopt the hexagonal ZrNiAl
structure.\cite{P13,P14,P15} A wealth of interesting magnetic properties has been
observed. CePdMg, CePtMg, and CeAuMg show long-range magnetic ordering below 2-3
K.\cite{P9} EuAgMg and EuAuMg order ferromagnetically with Curie temperatures of 22 and
37 K, respectively.\cite{P6} Also the Gd materials GdPdMg, GdPtMg, and GdAgMg are
ferromagnetic, with relatively high Curie temperatures of 96, 98, and 39 K,
respectively.\cite{P10} GdAuMg, on the other hand, orders antiferromagnetically, with a
N\'{e}el temperature of 81 K.\cite{P12}

Not much is presently known about the electronic structure of these materials. Basic
spectroscopic determination of the 4f valence of the rare earths is lacking, and it is
also unclear how band structure effects will play out for the $T$ $d$ bands in view of
the rather peculiar crystal structure. We have therefore set out to perform valence band
and core level photoelectron spectroscopy measurements. Here we chose for EuAuMg, GdAuMg,
YbAuMg and GdAgMg as model materials, since with the $d$ shell of the Au and Ag
constituents being closed, we can avoid complications which otherwise could occur as a
result of the intricate interplay between band formation and correlation effects usually
present in an open $T$ $d$ shell system. We also have calculated the band structure of
these materials using the local density approximation (LDA) and local density
approximation + Hubbard $U$ (LDA+$U$) scheme,\cite{Anisimov91} where the $U$ is used to
account for correlation effects in an open atomic-like $4f$ shell. As it turns out later,
it was necessary to include as well the spin-orbit coupling for the Au and Ag $d$ shells.

The starting rare earth, noble metal and magnesium materials were mixed in the ideal
1:1:1 atomic ratios and sealed in niobium or tantalum tubes under an argon pressure of
$\simeq 800$ mbar. These tubes were first rapidly heated to $\simeq 1370$ K and
subsequently annealed at $\simeq 870$ K for two hours. The samples were separated from
the tubes by mechanical fragmentation. No reaction with the metal tubes was observed. The
polycrystalline samples with dimensions of about $2$$\times 1$$\times 1$\,mm$^3$ are
light grey with metallic luster. All samples were pure phases on the level of X-ray
powder diffraction, and their transport and thermodynamic properties have been analyzed
recently.\cite{Hartmann05}

The photoemission spectra were recorded at room temperature in a spectrometer equipped
with a Scienta SES-100 electron energy analyzer and a Vacuum Generators twin crystal
monochromatized Al-$K_{\alpha}$ ($h\nu$ = 1486.6 eV) source. The overall energy
resolution was set to 0.4 eV, as determined using the Fermi cut-off of a Ag reference,
which was also taken as the zero of the binding energy scale. The base pressure in the
spectrometer was 1x10$^{-10}$ mbar, and the pressure raised to 2x10$^{-10}$ mbar during
the measurements due to the operation of the x-ray source. The samples were cleaved
\textit{in-situ} to obtain clean surfaces.

\begin{figure*}
     \includegraphics[width=0.80\textwidth]{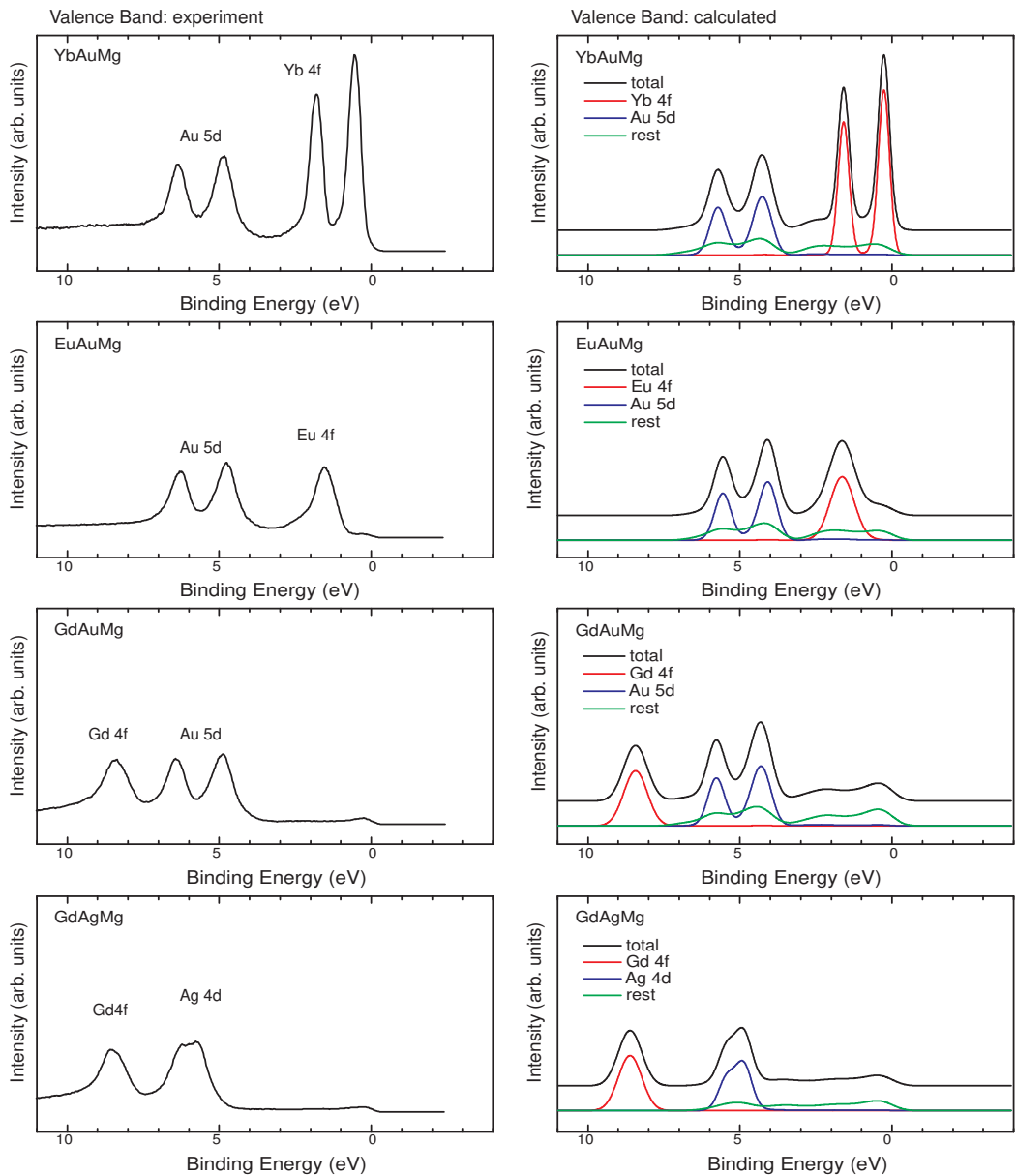}
          \vspace{-10mm}
     \caption{(color online) Experimental valence band photoemission
     spectra (left panels) and calculated valence band density of
     states (right panels) of polycrystalline YbAuMg, EuAuMg, GdAuMg,
     and GdAgMg (from top to bottom).}
    \label{spectra1}
\end{figure*}

The left panels of Fig. 1 show the valence band photoemission spectra of (from top to
bottom) YbAuMg, EuAuMg, GdAuMg and GdAgMg. A weak but clear cut-off at the Fermi level (0
eV binding energy) can be observed, consistent with these materials being
metals.\cite{Hartmann05} The three Au samples have in common that their valence bands
contain a two-peak structure at 5.0 and 6.4 eV binding energy. We therefore can assign
these two peaks to the Au $5d$. Consequently, the other peaks in the spectra can be
attributed to the rare-earth $4f$. YbAuMg shows a very sharp two-peak structure at 0.5
and 1.9 eV binding energy. The line shape as well as the energy separation of these two
peaks is characteristic for the Yb $4f_{7/2}$, $4f_{5/2}$ spin-orbit split $4f^{13}$
photoemission final states,\cite{Lang,Campagna,Wertheim} indicating that the Yb in the
ground state is $4f^{14}$, i.e. divalent, in agreement with the magnetic
susceptibility.\cite{Mishra01} The EuAuMg has a somewhat broader peak at 1.5 eV binding
energy, very typical for the multiplet structured $4f^{6}$ final states of divalent Eu
compounds and intermetallics having the high-spin $4f^{7}$ ground
state.\cite{Lang,Campagna} This assignment is supported also by the fact that the
integrated $4f$ intensity of the Eu sample is close to half of the Yb $4f$, i.e.
proportional to the number of $f$ electrons. The peak at 8.4 eV binding energy in the
GdAuMg sample is very similar in line shape and intensity to the 1.5 eV feature of the Eu
sample, indicating that this peak corresponds also to a $4f^{6}$ final state, and thus to
a ground state having the high-spin $4f^{7}$ configuration.\cite{Lang,Campagna} In other
words, the Gd is trivalent.

In comparing the two Gd samples, i.e. GdAuMg and GdAgMg, one can
clearly see that the line shape differs mainly in the spectral
range between 5 and 7 eV binding energy. We then can ascribe the
'flattened' peak at 6 eV in the GdAgMg to the Ag $4d$ band.
Remarkable is that the line shapes of both the Au $5d$ and the Ag
$4d$ bands in these materials are vastly different from those Au
$5d$ and Ag $4d$ bands of the elements, as shown in Fig. 2. The
noble metal $d$ bands in these rare-earth magnesium
intermetallics are so narrow that one can clearly observe the
splitting due to the spin-orbit interaction. The bands are much
less than 1 eV wide, i.e. about a factor of five smaller than in
the corresponding elements.

\begin{figure}
     \includegraphics[width=0.40\textwidth]{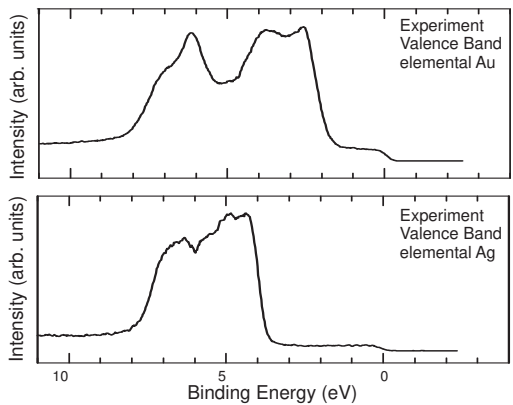}
     \caption{Experimental valence band photoemission
     spectra of elemental Au (top panel) and Ag (bottom panel).}
    \label{spectra2}
\end{figure}

To facilitate the interpretation of the experimental results, we have carried out band
structure calculations using the LDA and LDA+$U$ methods,\cite{Anisimov91} where the $U$
refers to the on-site Coulomb energy in an open $4f$ shell. We have also included the
spin-orbit coupling for the Au and Ag $d$ shells. Our calculations were performed by
using the full-potential augmented plane waves plus local orbital method.\cite{WIEN2k} We
took the crystal structure as determined by x-ray diffraction
measurements.\cite{P4,P10,P12} The muffin-tin sphere radii are chosen to be 3.2, 2.5,
2.6, and 2.7 Bohr for rare-earth, gold, silver, and magnesium atoms, respectively. The
cut-off energy of 16 Ryd is used for plane wave expansion of interstitial wave functions,
and 400 {\bf k} points for integrations over the Brillouin zone, both of which ensure
sufficient numerical accuracy. The spin-orbit coupling is included by the
second-variational method with scalar relativistic wave functions.\cite{WIEN2k} We
calculated the EuAuMg and GdAgMg in the ferromagnetic, the GdAuMg in the
antiferromagnetic and the YbAuMg in the paramagnetic state. The easy magnetization
direction is set along the c axis.

The right panels of Fig. 1 depict the results of the LDA and LDA+$U$ calculations. One
can clearly see that the density of states reproduces all the spectral features quite
well, including the very narrow and spin-orbit split bands of the Au $5d$ and the Ag
$4d$. Apparently, the crystal structure is such that the distance between the noble metal
atoms is sufficiently large as to make the overlap between the $d$ orbitals negligible.
The line shape of these narrow bands is in fact reminiscent of that of noble metal
impurities submerged in a low density $sp$ metal host.\cite{Weightman} The calculations
indeed indicate the presence of such an extremely broad $sp$ like band. This band, which
extends from 7 eV binding energy all the way to above the Fermi level, is labelled as
'rest' in Fig. 1, and consists of Mg $s$, Au/Ag $sp$ and $RE$ $spd$ bands. They in fact
build up the relevant states in the vicinity of the Fermi level. The contribution of the
$RE$ $4f$ bands to these is negligible, with the exception perhaps for the Yb system.

To address the issue concerning the $4f$ shell, we used the LDA method for the Yb system
in which we also include the spin-orbit interaction in the $4f$. The calculation finds
correctly the closed shell $4f^{14}$ configuration for the Yb, and the peak positions are
in good agreement with the experiment. For the Eu and Gd system, however, we have to
resort to the LDA+$U$ method, in which $U$ in the $4f$ shell is set to 6 eV for the Eu
and 7 eV for the Gd.  This is necessary to account for correlation effects in the open
atomic-like $4f$ shell: both Eu and Gd have the $4f^{7}$ configuration as found from
experiment and calculation. If $U$ were set to zero, the Eu and Gd $4f$ states would move
sizeably to the Fermi level, and the experimental peak positions would not have been
reproduced.

Despite the apparent successes of these band structure calculations, we also can observe
small deviations between theory and experiment. A closer look at the calculated positions
of the closed shell Au and Ag $d$ bands as well as of the closed shell Yb $4f$ orbitals,
reveals that they are too close to the Fermi level by several tenths of an eV. We
attribute such deviations as being caused by the inherent limitations of these mean-field
methods to calculate the dynamical response of a system.

\begin{figure*}
     \includegraphics[width=0.80\textwidth]{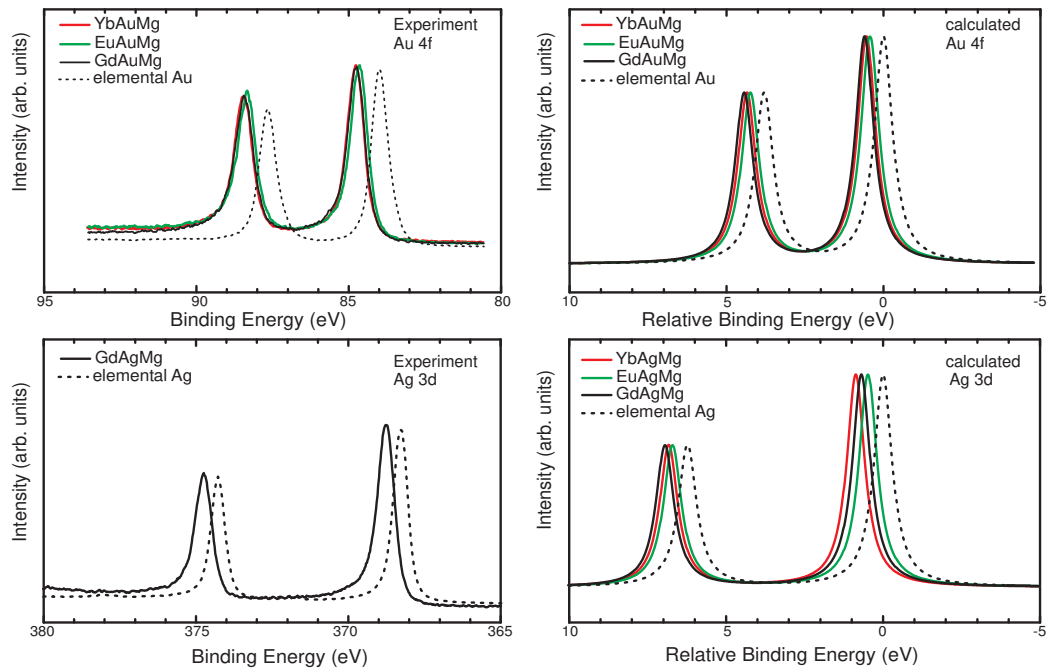}
     \vspace{-2mm}
     \caption{(color online) Left panels: experimental Au $4f$
     and Ag $3d$ core level photoemission spectra of
     YbAuMg, EuAuMg, GdAuMg, GdAgMg, and elemental Au and Ag.
     Right panels: corresponding simulated spectra based
     on the calculated Au $4f$ and Ag $3d$ energy positions.}
    \label{spectra3}
\end{figure*}

The left panels of Fig. 3 show the Au $4f$ and Ag $3d$ core
levels of the YbAuMg, EuAuMg, GdAuMg and GdAgMg samples (colored
solid lines). As reference we have also measured the corresponding
core levels in elemental Au and Ag (dashed lines). One can
clearly observe that the Au and Ag core levels in the
intermetallics have substantially higher binding energies than
those in the elements, i.e. $\approx$0.7 eV for the Au $4f$ and
$\approx$0.5 eV for the Ag $3d$. This can be taken as an
indication that the charge density at the Au and Ag sites in the
intermetallics is appreciably lower than in the elements. LDA
band structure calculations confirm this picture. We have
calculated that the energies of the Au and Ag core levels of the
intermetallics are indeed lower (i.e. at higher binding energy)
than in the elements, and this is shown in the right panels of
Fig. 3 where we have simulated the spectra based on these
calculated energy positions. These findings are consistent with
the crystal structure and the very narrow band width of the $d$
bands: the Au and Ag atoms in the intermetallics could be
considered as impurities submerged in a low density $sp$ metal
host.

To conclude, we have investigated the electronic structure of the
equiatomic EuAuMg, GdAuMg, YbAuMg and GdAgMg intermetallics by
x-ray photoelectron spectroscopy and LDA/LDA+$U$ band structure
calculations. We found a closed shell $4f$ configuration for the
Yb ('divalent') and a $4f^{7}$ for the Eu ('divalent') and Gd
('trivalent'). The states in the vicinity of the Fermi level are
given by a broad $sp$ like band consisting of Mg $s$, Au/Ag $sp$
and $RE$ $spd$ states. The contribution of the $RE$ $4f$ is
negligible, except perhaps for the Yb system. We also found that
the Au and Ag $d$ bands are extraordinarily narrow, as if the
noble metal atoms were impurities submerged in a low density $sp$
metal host. This calls for a follow-up study of the electronic
structure of equiatomic intermetallics in which the noble-metal
atoms are replaced by transition-metal atoms with an open $d$
shell, since in view of the extremely narrow band width even
modest electron correlation effects could already play an
important role for the properties.

We acknowledge Lucie Hamdan for her skillful technical assistance.
We thank the Degussa-H\"{u}ls AG for the generous gift of noble
metals. This work was supported by the Deutsche
Forschungsgemeinschaft through the priority programme SPP 1166
"Lanthanoidspezifische Funktionalit\"{a}ten in Molek\"{u}l und
Material".

\end{document}